\documentclass[journal]{IEEEtran}
\IEEEoverridecommandlockouts
\usepackage{short_commands_article_IEEE}

\def\BibTeX{{\rm B\kern-.05em{\sc i\kern-.025em b}\kern-.08em
    T\kern-.1667em\lower.7ex\hbox{E}\kern-.125emX}}
\begin{document}
\bibliographystyle{IEEEtran}  

\title{Model-Based Data-Efficient and Robust Reinforcement Learning}

\author{Ludvig Svedlund, Constantin Cronrath, Jonas Fredriksson, and Bengt Lennartson~\IEEEmembership{Fellow, IEEE}
\thanks{Ludvig Svedlund, Constantin Cronrath, Jonas Fredriksson, and Bengt Lennartson are all with the Division of Systems and Control, Department of Electrical Engineering, Chalmers University of Technology, Gothenburg, Sweden (e-mail: ludvige@chalmers.se).}
}
\maketitle
\begin{abstract}
A data-efficient learning-based control design method is proposed in this paper. It is based on learning a system dynamics model that is then leveraged in a two-level procedure. On the higher level, a simple but powerful optimization procedure is performed such that, for example, energy consumption in a vehicle can be reduced when hard state and action constraints are also introduced. Load disturbances and model errors are compensated for by a feedback controller on the lower level. In that regard, we briefly examine the robustness of both model-free and model-based learning approaches, and it is shown that the model-free approach greatly suffers from the inclusion of unmodeled dynamics. In evaluating the proposed method, it is assumed that a path is given, while the velocity and acceleration can be modified such that energy is saved, while still keeping speed limits and completion time. Compared with two well-known actor-critic reinforcement learning strategies, the suggested learning-based approach saves more energy and reduces the number of evaluated time steps by a factor of 100 or more.

{\em Note to practitioners---}
Optimizing the energy consumption of moving devices such as robots, automated guided vehicles, or trucks is a common task in automation and logistics. While the path along which the device is traveling might have been decided by, for instance, the factory layout, the velocity and acceleration of the device on this path may affect its energy consumption. Complicating factors in this problem may be that system dynamics are not fully known, data is limited, and deadlines must be met. To address these issues, we propose a learning-based control design method that efficiently estimates parameters of a simple model from physical knowledge, which is then used to 1) optimally plan a reference velocity and 2) compute feedforward control signals. Paired with a feedback controller, this leads to robust performance with respect to the desired control objectives even when some system dynamics were neglected. 

In practice, this greatly simplifies the consideration of state, action, and time constraints and proves to be significantly more data-efficient than deep reinforcement learning. In future work, we will generalize the method to other system model classes and optimization problems.
\end{abstract}


\section{Introduction}
\IEEEPARstart{K}{nowledge} about the dynamic behavior of a system is crucial in the design and optimization of feedback control systems. This knowledge is often based on system models, but also data from physical experiments. Such experiments can be step responses or time series analysis, where the system is excited by some type of random input signals, often in combination with closed-loop control \cite{ljung:1999}. The controller is then adapted such that the behavior of the closed-loop system is improved, naturally based on an online optimization procedure.

Reinforcement learning (RL) is a popular example of such an adaptive control strategy. A system model is then estimated, and based on this model, a controller is designed. Alternatively, a controller is directly determined such that an estimated criterion is optimized. The first version is called model-based RL, while the second one is called model-free RL \cite{sutton2018reinforcement, bertsekas2019reinforcement, bds:2010}. A special form of the model-free version is a state feedback controller that is optimized given a traditional linear quadratic criterion, formulated among other things in \cite{bradtke1994adaptive,lewisRL:2012}. 

Many robustness aspects of RL have been studied. In~\cite{moos2022robust}, the literature on robustness in RL is subdivided into approaches that target 1) robustness against uncertainties in the system dynamics, 2) robustness against external disturbances, 3) robustness against control input disturbances, and 4) robustness against sensor disturbances. One type of uncertainty in system dynamics is simplified or neglected dynamics. In principle, all dynamic models of physical systems are to some extent simplified versions of the actual dynamics.
The most common examples are short time constants and resonances that often can be neglected due to their high-frequency behavior. The sensitivity to such neglected high-frequency dynamics is systematically investigated in \cite{ludvig:2025}, and the results show that especially model-free RL is more sensitive than expected to such unmodeled dynamics. Even when small time constants or high-frequency resonances are involved in the system, but not included as states in the RL strategy, the closed-loop performance must be significantly detuned to preserve stability of the closed-loop system. A short summary of these findings is given in the first part of this paper, where the conclusion is that a model-based RL strategy is preferable.

Two more well-known challenges with model-free RL are data efficiency and formulation of the reward function. In practice, the control objective is often a mix of desired optimal behavior and hard constraints, both related to states and control signals. A constraint particularly important in the automation and operations research context is a constraint on the final completion time of a task. Such constraints are easily expressed in mathematical (optimization) programs. In model-free RL, however, they have to be translated to weighted terms in the reward function. The weighting of multiple objectives and hard constraints is a major design challenge, and a misconfiguration can lead to reward gaming~\cite{amodei2016concrete}. This may result in unexpected and undesired\,--\,even unsafe\,--\,\textit{``optimal''} control behaviors that may not immediately be distinguishable from the transient performance during learning. Several extensions to RL have thus been proposed, such as~\cite{altman1999constrained,lennartson2020reinforcement_WODES,liu2021policy,cronrath2022relevant}.

Data efficiency is a fundamental problem, since estimation and iterative optimization are integrated in model-free RL. A new criterion estimation must be performed in every optimization iteration. This is significantly simplified in model-based RL, where only one model is estimated. Most prominent model-based RL methods in the RL community try to solve this dual control problem as well, i.e. they try to learn the model while learning an optimal controller, and they do so in an iterative fashion. For instance, Sutton's DynaQ adds new transitions to the model and re-learns the $Q$-function with this model in every learning iteration. This is confirmed in \cite{tu2019gap}, where the sample complexity of model-free and model-based methods is studied. The model-based methods for policy evaluation and optimization use sampled trajectory data to estimate the system dynamics, which is shown to be at least a factor of the state dimension more sample-efficient than the model-free methods.

Motivated by robustness, data efficiency, and optimization constraints, a specific model-based RL strategy is proposed in this paper. This strategy also has the benefit that the learning and optimization are more open, flexible, and modular, and it is therefore called {\em modularized RL}. First, a system model is estimated and evaluated, where prior knowledge of expected behavior can be used, and different model structures can be chosen depending on the system behavior, either input-output or state-space models, linear or nonlinear, gray-box or black-box models \cite{ljung:1999,ljungGlad:2021}.

Based on the estimated model, a controller is designed and optimized. It can be based on several linearized models or a nonlinear design method. Different types of robust control strategies can be used, including even simple PI and PID controllers. It can also be separated into a nonlinear feedforward control function and local linear feedback controllers, as proposed in this paper. In our combined feedforward/feedback control system, the reference input signal is optimized to keep specific constraints. When the path is known, moving from one position to another, this optimization can be performed by a simple optimization strategy called {\em temporal optimization}, first applied in a simplified version in energy optimization of robots \cite{riazi2017energy} and for speed profile planning in \cite{liu2017speed}. 

In several applications, the path is given or a few alternatives are considered, such as transportation on roads and in factories, as well as in robot stations. In robot stations, the path is first decided to avoid collisions, and then velocities, accelerations, and waiting states are decided to keep the desired make span and optionally minimize energy consumption. Indeed, the separation between path planning and time optimization (scheduling) is a common strategy in many applications. Thus, considering a given path is often a reasonable assumption, and then temporal optimization is both a flexible and efficient optimization strategy.

The proposed modularized RL strategy is evaluated on eco-driving, where the speed of a truck and a vehicle is adjusted to save energy, while still keeping speed limits and desired completion times. The proposed optimization procedure is compared with some well-known actor-critic RL strategies, and the evaluation confirms the significant improvement using our proposed method when data efficiency and the design of reward functions are evaluated.

To summarize the main contributions of this paper, it highlights the fact that traditional model-free RL can be very sensitive to unmodeled dynamics. A modularized model-free RL strategy is then proposed, based on estimation of a nonlinear state space model, a combined nonlinear feedforward and feedback function, followed by an optimization strategy called temporal optimization that is very flexible in terms of hard state and input constraints. The modularized RL strategy is also shown to give good and data-efficient results in two electric vehicle applications where energy is minimized.

In the next section, model-free RL is briefly explained, followed by model-based RL in Section 3 and some robustness aspects in Section 4. In Section 5 the proposed modularized RL strategy is presented, including temporal optimization in Section 6. Two evaluations on electric vehicles are presented in Section 7 followed by some conclusions and topics for further studies in Section 8.

\section{Model-free Reinforcement Learning}\label{sec:RL}
The basic principles of model-free RL are presented in this section. First, the special case of a linear quadratic (LQ) optimization criterion is presented, followed by a short summary of deep RL based on neural networks, generally applied to non-linear systems. 

\subsection{Linear Quadratic Reinforcement Learning}\label{ssec:LQRL}
Consider a discrete-time system with quadratic stage cost 
\beq
\rho(x_k,u_k) = x_k^TQ_x x_k + u_k^TQ_u u_k,
\eeq{eq:rxu}
where the weighting matrices $Q_x$ and $Q_u$ are symmetric. For a state feedback control policy  $u_k=\mu(x_k)$, an infinite horizon cost function can be formulated as
\[
V^\mu(x_k)\!=\! \sum_{i=k}^{\infty} \rho(x_i,\mu(x_i))\smm=\smm \rho(x_k,\mu(x_k))+ V^\mu(x_{k+1})
\] 
The rewritten recursive formulation is called Bellman's equation \cite{bertsekas2019reinforcement}. 

\subsubsection{Quadratic Q-function}
To obtain an optimal control policy, Bellman's equation is reformulated by introducing the $Q$-function 
\bea
Q(x_k,u_k) &\hspace{-1.5ex}=\hspace{-1.5ex}& \rho(x_k,u_k)+ V^\mu(x_{k+1}) \yspa{1.2ex}\nonumber\\
&\hspace{-1.5ex}=\hspace{-1.5ex}& \rho(x_k,u_k)+ Q(x_{k+1},\mu(x_{k+1}))
\eea{eq:Qupdate}
where we observe that $V^\mu(x_k) = Q(x_k,\mu(x_k))$. 
When the stage cost is quadratic as in \rf{eq:rxu}, both the value function $V^\mu$ and the $Q$-function will have a quadratic form. Thus, the $Q$-function can be expressed as
\beq
Q(x,u,\theta)=\mat{\, x^T \!\!&\!\! u^T \,} \smm \mat{\, S_{xx}(\theta) \!\!&\!\! S_{xu}(\theta) \, \yspa{1.2ex}\cr\,  S_{xu}^T(\theta) \!\!&\!\! S_{uu}(\theta) \,}  \smm\sm\mat{\, x \, \yspa{1.2ex}\cr\, u\,} =\vphi^T\smm(x,u)\saa\theta
\eeq{eq:Qtheta}
For a scalar system with one state $x_k$ and one control signal $u_k$, the $S$-matrices are parameterized as
$S_{xx}(\theta)= \theta_1$,  $S_{xu}(\theta)=\theta_2$, and $S_{uu}(\theta) = \theta_3$, and the regression vector $
\vphi^T(x,u)=\mat{\, x^2 \smm&\smm  2xu \smm&\smm u^2\,}$. For the general case, see \cite{lewisRL:2012}.

\subsubsection{Policy iteration}
The optimal control policy $\mu^*\smm(x_k)$ is achieved by selecting the parameter vector $\theta$ such that minimization of $Q(x_k,u_k,\theta)$ generates this optimal policy. This can be achieved by policy iteration \cite{bertsekas2019reinforcement} where a given parameter vector $\theta_{\ell-1}$ and related policy $\mu_{\ell-1}(x_k)$ are updated by minimizing the temporal difference (TD) error for $Q$ with respect to $\theta$, \cf \rf{eq:Qupdate} and  \cite{sutton2018reinforcement}. The minimization is here formulated by introducing the regression model
\bes
\rho(x_k,u_k) &\hspace{-1.5ex}=\hspace{-1.5ex}&  Q(x_{k},u_k,\theta) - Q(x_{k+1},\mu_{\ell-1}(x_{k+1}),\theta) + \veps_k \yspa{1.2ex}\\
&\hspace{-1.5ex}=\hspace{-1.5ex}& \lp\vphi^T(x_k,u_k)-\vphi^T(x_{k+1},\mu_{\ell-1}(x_{k+1}))\rp\theta +  \veps_k 
\ees
Given data ($x_k,u_k,x_{k+1}$) from $k=1$ to $N$, and solving the related normal equation for this regression model \cite{ljung:1999}, the sum of the squared TD-errors $\sum_{k=1}^N\veps_k^2$ is minimized. The resulting $\theta_\ell$ generates the updated $Q(x,u,\theta_\ell)$ in the policy iteration.

\subsubsection{Policy improvement} 
Given the updated $\theta_\ell$, the policy is  improved as
$
\mu_{\ell}(x) = \text{arg}\min_{u}Q(x,u,\theta_\ell).
$
This minimum is obtained when the gradient
$
\frac{\partial Q(x,u,\theta_\ell)}{\partial u}=0.
$
Performing this gradient condition on \rf{eq:Qtheta} results in the matrix equation $2S^T_{xu}(\theta_{\ell})x+2S_{uu}(\theta_{\ell})=0$, and the improved state feedback policy 
\beq
u=\mu_{\ell}(x)=-K_{\ell}x,  \qquad K_{\ell}=S_{uu}^{-1}(\theta_{\ell})S^T_{xu}(\theta_{\ell}).
\eeq{eq:SFB}
Performing this policy iteration with a stabilizing initial control policy $u=K_0x$, and iterating until $K_{\ell}\approx K_{\ell-1}$, results in the optimal control policy. 

\subsubsection{Model-free strategy} This computation does not include any specific model knowledge, only state and input information $x_k,u_k,x_{k+1}$ from $k=1$ to $N$, and an initial state feedback gain. For stable systems, this can be $K_0=0$, and the convergence of this iteration is shown in~\cite{bradtke1994adaptive}. When the state vector is updated by the state-space model $x_{k+1}=Ax_k+Bu_k$, this data-based strategy is an interesting alternative to the traditional LQ state feedback control solution, obtained by solving a Riccati equation \cite{anderson:2014}.

LQRL is well known in the control community, formulated among others in \cite{bradtke1994adaptive,lewisRL:2012}. The minimalist presentation in this subsection is included to illustrate the basic principles of model-free RL compared to model-based RL. Furthermore, the robustness of both strategies for linear systems with additional but neglected high-frequency dynamics is evaluated in \secr{sec:robust}.

\subsection{Deep Reinforcement Learning}
Model-free RL for nonlinear state-space systems is mostly formulated using function approximations. In deep reinforcement learning (DRL), the function approximation is based on deep neural networks (NNs) \cite{haykin:1998, goodfellow:2016}. One of the most common DRL strategies is actor-critic RL, \cf \cite{sutton2018reinforcement} Ch.~13, where policy iteration includes a NN  for the $Q$-function and another NN for the control policy improvement. The update of the $Q$-function is called the {\em critic part}, and the improvement of the control policy is called the {\em actor part}. In the linear quadratic version presented above, the critic part corresponds to the estimation of the $S$-matrices in the $Q$-function in \rf{eq:Qtheta}, while the actor part is obtained analytically by \rf{eq:SFB}.

\subsubsection{TD3 and SAC}
The most popular actor-critic DRL algorithms are available and presented at {\em OpenAI Spinning Up} \cite{OpenAISpinningRL}, which also gives an excellent introduction to DRL. One prominent DRL method is Twin Delayed DDPG (TD3) \cite{fujimoto2018addressing}, and another one is Soft actor-critic (SAC) \cite{haarnoja2018soft}. Both methods estimate 
\[
\mu_{\ell}(x) = \text{arg}\min_{u}Q(x,u,\theta_\ell)
\]
directly through a neural network instead of computing the minimizer of $Q(x,u,\theta_\ell)$ analytically. This is generally of computational advantage in cases with multiple continuous control inputs. TD3 estimates a deterministic policy, whereas SAC uses a stochastic policy. The policy is then improved by taking gradient steps on 
\[
Q(x,\mu_{\ell}(x, \vartheta_\ell),\theta_\ell)
\]
with respect to the parameters $\vartheta_\ell$ of the policy estimator. Clearly, a good estimate of 
$
Q(x,u,\theta_\ell)
$
is crucial to the success of this policy improvement step. 

In both methods, the $Q$-function is approximated through minimization of the squared TD-errors with respect to $\theta_\ell$. To reduce instabilities in this process, both TD3 and SAC maintain two estimates of the $Q$-function, but only use the more conservative estimate in computing the $Q$-value in the next state $x_{k+1}$. Furthermore, TD3 adds noise to the action selected in the next state, while SAC naturally samples an action from its stochastic policy. Lastly, the Q-values of the next states are computed from a running average of past $\theta_\ell$ and, in the case of TD3, past $\vartheta_\ell$. All three aspects of these two methods add to the robustness of the learning procedure by reducing the risk of overfitting the actor or critic NN on approximation errors.

\section{Model-based Reinforcement Learning}\label{sec:MBRL}
Parameters in the optimization criterion are estimated in model-free RL, \cf \rf{eq:Qtheta}. Actor-critic RL includes an additional set of parameters that are estimated in the control policy improvement. In model-based RL, on the other hand, a model of the system to be controlled is estimated.

Compared to the policy iteration in model-free RL, where new parameter vectors $\theta_\ell$ are estimated in every iteration $\ell$, only one iteration is required where parameters are estimated in model-based RL. This procedure is also more open, flexible, and modular in the sense that first the model can be evaluated, to see if it is reasonable or not. Then, a controller is designed, including evaluation of stability margins and performance in an offline study, before the designed controller is evaluated on the real system. These aspects are further developed in the following sections.

Some alternative model estimation strategies are introduced in the following two subsections, including both linear and nonlinear model structures. 

\subsection{Linear model estimation}
Linear discrete-time models can be formulated as state-space models, but also using polynomials in the time shift operator $q$,  meaning that $q^{-1}y_k = y_{k-1}$ \cite{astrWittCCS:1997}.

\subsubsection{ARX models}
An autoregressive (AR) model including a polynomial $\mc A(q)$ operating on an output signal $y_k$ and an unknown disturbance $e_k$, together with a polynomial $\mc B(q)$, operating on a known eXternal input $u_k$, is called an ARX model. An ARX model of order $n$ is therefore formulated as
\[
\mc A(q)y_k=\mc B(q)u_k+e_k
\]
where $\mc A(q)=1+a_1q^{-1}+\ldots a_nq^{-n}$ and $\mc B(q)=b_1q^{-1}+\ldots b_nq^{-n}$. By reformulating this model as a regression model 
\[
y_k=\vphi^T_k\theta+e_k
\]
with $\vphi_k = \mat{\, -y_{k-1}  \,\,\, \cdots \,\,\, -y_{k-n} \!&\! u_{k-1} \,\,\, \cdots \,\,\, u_{k-n} \,}^T$ and $\theta\in \Re^{2n}$, the estimated parameters in $\what\theta$ are obtained by solving the related normal equation for this regression model \cite{ljung:1999}. This model is easily extended to multiple inputs \mbox{$(n_u\geq 1)$} and multiple outputs $(n_y\geq 1)$ by generalizing $\mc A(q)$ to a $n_y\times n_y$ polynomial matrix  and $\mc B(q)$ to a corresponding $n_y\times n_u$ matrix. 

\subsubsection{Linear state-space model}
When all states in a linear state-space model
\begin{align}
\begin{split}
x_{k+1} =& Ax_k + Bu_k \yspa{1.2ex} \\
y_{k}=& Cx_k + Du_k \label{eq:ss}
\end{split}
\end{align}
are measured, \ie $y_k=x_k$ and hence $C=I_{n_x}$ and $D=0$, the parameters in the $A$ and $B$ matrices can be estimated by formulating an ARX model with $n_y=n_x$ output signals and first order polynomials in both $\mc A(q)$ and $\mc B(q)$, that is $n=1$. 

\subsubsection{State feedback control design}
A state feedback controller can now be computed by estimating such an ARX model with $n_x$ outputs and first-order polynomials ($n=1$). The resulting state-space model $(A,B)$, together with the same LQ criterion as in model-free RL with stage cost matrices $Q_x$ and $Q_u$, generates the LQ state feedback gain $K$  by solving a Riccati equation based on these matrices \cite{anderson:2014}. 

\subsection{Nonlinear model estimation} \label{ssec:identNL}
To handle nonlinear systems, two different nonlinear model structures are recommended, depending on the know\-ledge of the system to be controlled.

\subsubsection{Nonlinear black-box models}
The ARX model is a black-box model where no physical knowledge is introduced in the estimation procedure. The required number of parameters $n$ in the model can be selected by evaluating the estimated model. This is done by comparing the simulated output from the estimated model with the real output signal, complemented by various statistical measures on the model accuracy~\cite{ljung:1999}. 

A natural generalization of the black-box ARX model to nonlinear systems is to add function approximations to the ARX model structure. Common examples are (deep) neural networks and wavelets \cite{ljung:1999,ljungGlad:2021}. In MATLAB's System Identification Toolbox \cite{ljung:2025}, such a nonlinear extension is obtained by only changing the function call from {\tt arx} to {\tt nlarx} and adding the desired function approximation. To get good results with this nonlinear model estimation strategy, the input signal excitation must involve the whole signal range where the model is expected to be used.

\subsubsection{Nonlinear gray-box models}
When some knowledge of the system to be controlled is available, a gray-box model is an interesting alternative model structure. Physical knowledge of dynamical systems is mainly formulated as continuous-time models, often in terms of differential equations, such as Newton's law for moving devices as well as Kirchhoff's first and second laws for electric circuits. The nonlinear model in \secr{sec:modRL} has a well-known structure, where the wind resistance causes a damping force that is proportional to the square of the speed. It means that the model structure can be assumed to be known, while specific parameter values in the dynamic model are still requested. 

A continuous-time non-linear state-space model
\begin{align}
\begin{split}
\dot x = f(x,u,\theta)
 \yspa{1.2ex} \\
y = g(x,u,\theta) \label{eq:ss}
\end{split}
\end{align}
is therefore introduced, where unknown parameters in the parameter vector $\theta$ are also included in the model, together with inputs $u$ and outputs $y$ that are assumed to be known at discrete-time instances $t_k=kh$ where $h$ is the sampling interval. Ones again, MATLAB's System Identification Toolbox provides a function {\tt idnlgrey} \cite{ljung:2025} by which the parameter vector $\theta$ in \rf{eq:ss} is estimated, based on discrete-time input/ouput data and a function which defines how parameters in the $\theta$ vector and the variables in the vectors $x$, $y$, and $u$ are related to each other. 

\section{Robust Reinforcement Learning} \label{sec:robust}
Based on a fair evaluation method for closed-loop system performance and stability margins, the sensitivity to neglected dynamics in RL will now be illustrated by a simple example. 
In every mechatronic system, including a servo motor model for a rotating system, there are always added dynamics in terms of additional time constants and resonances. One neglected time constant is the electrical time constant in the motor. Thus, consider the transfer function
\[
G(s)=\frac{1}{s(1+s)(1+\tau s)}
\]
including integral action from rotation speed to the output angle, a normalized time constant equal to one, and an additional time constant $\tau$. Let $x_1$ be the state of the output signal, the angle, and $x_2$ the angular velocity, while $x_3$ is the additional state, representing the time constant $\tau$. When $\tau=0$, $x_3=u$, meaning that this state can be removed. 

The question is now how small values of $\tau$ influence the closed-loop behavior when model-free and model-based RL are used to design a state feedback controller 
\beq
u=-Kx+K_rr,
\eeq{eq:uk}
The stage cost is $\rho(x,u)=x_1^2+Q_u u^2$. In the model-free case, the linear quadratic procedure in \secr{sec:RL} is applied, and in the model-based version, a second-order linear discrete-time state-space model is estimated according to \secr{sec:MBRL}. Based on this model, the Riccati equation for the linear quadratic criterion is solved. The sampling period $h=0.1$, and in both cases, there is no feedback from the third state, since this state (mode) is neglected. 

\subsubsection{Closed loop performance measure}
Now, introduce the notation $(A,B,C,D)$ for a generic linear discrete-time state space model. Including the control signal \rf{eq:uk} in the system model $(A,B,C,0)$, the state-space model from the reference signal $r$ to the output $y$ can then be expressed as 
\[
G_{ry}=(A-BK,BK_r,C,0).
\]
As a performance measure for this closed-loop system $G_{ry}$, we use the rise time, defined as the time it takes the output signal $y$ in a step response to increase from $10\%$ to $90\%$ of the final value $y(\infty)$.

\subsubsection{Stability margins}
The loop transfer $L$ can be expressed as $(A,B,K,0)$, which for one control signal generates the frequency function $L(e^{j\omega h})$. The sensitivity functions $S$ and $T$ may then be expressed by the transfer functions
\[
S=\frac{1}{1+L} \quad \text{and} \quad T=1-S=\frac{L}{1+L}
\]
Now, introduce the following constraints on the maximum values of $S$ and $T$ in the frequency domain
\bes
M_S &\hspace{-1.5ex}=\hspace{-1.5ex}& \max_{\omega}S(e^{j\omega h}) \leq 1.7 \\
M_T &\hspace{-1.5ex}=\hspace{-1.5ex}& \max_{\omega}T(e^{j\omega h}) \leq 1.3.
\ees
$M_S$ and $M_T$ are common measures of the stability margin, where $1/M_S$ is the minimal distance between the loop $L(e^{j\omega h})$ and the point $(-1,0)$ in the Nyquist curve. The constraint on $M_S$  means that the distance is larger or equal to $1/1.7=0.59$. The additional constraint on the complementary sensitivity function $M_T$ reduces overshoots in the closed loop response, especially for systems $(A,B,C,0)$ including integral action and/or unstable poles.

\subsubsection{Evaluation by constraint optimization}
To summarize, the evaluation is performed by searching for 
\[
\min t_r \quad \text{subject to} \quad M_S\leq 1.7, \; M_T\leq 1.3
\]
This constraint optimization problem has only the control penalty $Q_u$ as a decision variable. Starting with a high value and reducing $Q_u$ until at least one of the constraints on $M_S$ and $M_T$ is reached gives the minimal rice time $t_r$.

Solving this constraint optimization problem for different values of the neglected time constant $\tau$ results in rise times for model-free and model-based RL according to \tabr{tab:extraTimeConst}

\begin{table}[h]
    \centering
    \caption{Minimal rise time for model-free RL $t_r^{mf}$ and model-based RL $t_r^{mb}$ for various extra but neglected time constants $\tau$.\yspa{1.2ex}}
    \label{tab:extraTimeConst}
    \begin{tabular}{c|cccccc}
    \toprule
     $\tau$    & 0.2 & 0.1 & 0.05 & 0.02 & 0.01 & 0   \\
    \midrule
    $t_r^{mf}$ & 71.9 & 54.2 & 23.0 & 5.14 & 2.97 & 0.22 \\
    $t_r^{mb}$ & 2.15 & 1.11 & 0.64 & 0.46 & 0.31 & 0.22 \\
    \bottomrule
    \end{tabular}
\end{table}

The added dynamics means that the feedback control law must be significantly detuned by increasing the control loss $Q_u$, to keep the demanded stability margins.  This required detuning results in a dramatic increase in the rise time $t_r$, especially for the model-free RL strategy. Even for quite small~$\tau$, the increase in $t_r$ is significant to be able to stabilize the closed loop system. 

This sensitivity analysis of neglected small time constants as well as high-frequency resonances is presented in more detail in  \cite{ludvig:2025}. In this paper, it is also shown that adding low-pass filtering of the states and control signals reduces the sensitivity to this type of neglected high-frequency dynamics significantly. Still, the model-free RL strategy is less robust due to instability problems not only for smaller control penalties $Q_u$, but also for higher values and stable solutions in between. 

For model-based RL, filtering is not critical, but more significant neglected time constants also require detuning by increasing $Q_u$ to keep required stability margins. From a robustness point of view, the conclusion is, however, that model-based RL is preferable compared to model-free RL.


\section{Modularized Reinforcement Learning}\label{sec:modRL}
As already discussed in \secr{sec:MBRL}, model-based RL is a modular strategy in the sense that system model estimation and optimization of a feedback control function can be developed and evaluated individually and step by step. This is done before the final learned and optimized controller is evaluated on the real system. 

A more specific modular RL strategy is presented in the following two sections. It includes system model estimation and a combined feedback/feedforward strategy, assuming an arbitrary reference signal as input to the controlled system. This is followed by an optimal reference signal generation based on an efficient constrained optimization method called temporal optimization. The proposed modular RL strategy is focused on time and energy optimization of moving devices, but the principles are generic and can also be applied in other application areas.

\subsection{Estimation of a nonlinear dynamic model}
The dynamics of a moving device can be formulated as a force or torque balance 
\beq
u = \pi(v,\dot v, \alpha,\theta)
\eeq{eq:uBalance}
where the control signal $u$ depends on the velocity $v$, the acceleration $\dot v$, and optionally the slope $\alpha$, as well as a set of unknown parameters summarized in a parameter vector~$\theta$. The slope, for instance, on a road, is included as a typical load disturbance. It can be included as a known disturbance, which can be obtained by map and GPS information. Alternatively, it is unknown and then compensated by a feedback control loop.
To be able to determine the parameters in $\theta$ by experimental data, measuring $v$ and $u$, and optionally $\alpha$, the relation between these time-varying variables is assumed to be expressed by the nonlinear differential equation
\beq
\dot v = \theta_1 u +\theta_2  +\theta_3 v + \theta_4 v^2 +\theta_5 \alpha + \theta_6 \alpha^2 
\eeq{eq:NLdiff}
This is close to a black-box model, where second-order polynomials are introduced for the dependency on the velocity~$v$ and the slope $\alpha$, while a linear relation between the acceleration $\dot v$ and the control signal $u$ (force or torque) is motivated by Newton's law. 

The unknown parameters  $\theta_1,\ldots,\theta_6$ in the nonlinear differential equation \rf{eq:NLdiff} can be estimated by the {\em prediction error method} \cite{ljung:1999}, given the velocity $v_k$, slope $\alpha_k$ and control input $u_k$ for a number of discrete-time instances $t_k=kh$, $k=0,\ldots,N$, where $h$ is the sampling interval.  As mentioned in \secr{ssec:identNL}, this estimation method is specifically implemented for continuous-time state-space models in the function {\tt idnlgrey} in MATLAB's System Identification Toolbox~\cite{ljung:2025}.

With minor additional but neglected dynamics and moderate noise levels in available data, this estimation procedure generates an estimated parameter vector $\what\theta$ with required accuracy, as will be shown in \secr{sec:eval}. If neglected states have more significant dynamics, as discussed in \secr{sec:robust}, the parameter accuracy is reduced. Then either additional states must be included in the estimated state-space model or a higher order nonlinear black-box model, such as NLARX, also mentioned in  \secr{ssec:identNL}, can be used. Since the resulting dynamic model is then a discrete-time model, additional complications appear that will be discussed at the end of the paper.

\subsection{Data-based feedforward control}
A desired reference velocity $v^r$, the slope~$\alpha$, the estimated parameter vector $\what\theta$, and an Euler approximation of the reference acceleration $\dot v^r_k\approx(v^r_{k+1}-v^r_k)/(t_{k+1}-t_k)$, generate with the balance equation \rf{eq:uBalance} a feedforward control signal
\beq
u^r = \pi(v^r,\dot v^r, \alpha, \what\theta).
\eeq{eq:uref}
Assume now that this balance equation, for the estimated parameters in $\what\theta$, correctly describes the dynamic behavior of the real physical system. Applying the control signal $u=u^r$ and assuming $\what\theta=\theta$ then implies that
\[
\pi(v,\dot v, \alpha,\theta) = \pi(v^r,\dot v^r, \alpha,\what\theta)
\] 
Except for pathological examples, this implies that the real velocity is equal to the desired reference velocity, \ie \text{$v=v^r$}. Of course, the assumption above is unrealistic, but the feedforward control signal $u^r$ can still be expected to generate a system velocity $v$ that is quite close to the desired reference velocity $v^r$.  Additional improvements are also achieved by adding a feedback control loop.

\subsection{Velocity Dependent Feedback Control} \label{ss:FB}
Additional disturbances and modeling uncertainties are naturally compensated by adding a classical {\em feedback control loop} to the feedforward control strategy.  Since most feedback design methods are based on linear models, the nonlinear model is linearized.

Linearization of the nonlinear differential equation \rf{eq:NLdiff} around the reference velocity $v^r$ and the 
feedforward control signal $u^r$ gives, for $\Delta v=v-v^r$ and $\Delta u=u - u^r$, the linear state space model
\[
\Delta \dot v = a(v^r)\Delta v+b\Delta u,
\]
where $a(v^r)=\theta_3+2\theta_4 v^r$ and $b=\theta_1$. A piecewise constant control signal that is updated after each sampling interval $h=t_{k+1}-t_k$, then results in the discrete-time model
\[
\Delta v_{k+1} = \alpha(v^r)\Delta v(t_{k})+\frac{b}{a(v^r)}(1-\alpha(v^r)) \Delta u_k,
\]
where $\alpha(v^r)=\exp(-a(v^r)h)$. For a system with multiple states, a Taylor series expansion generates the corresponding linearized state-space model \cite{ljungGlad:2021}.

\parag{Linear quadratic optimal feedback control} Any suitable linear control design method can be selected based on the linearized model. To follow up on the model-free optimization criterion in \secr{sec:RL}, we also choose a state feedback controller that minimizes a linear quadratic criterion.  To be able to compensate load disturbances, for instance varying road slopes in \rf{eq:NLdiff}, integral action is included in the controller. Thus, introduce the integral state $\Delta v_{I k}= \sum_{\ell=0}^{k-1} \Delta v_\ell$
and the state vector $x=\mat{\Delta v &\Delta v_{I}}^T$. Minimizing the linear quadratic criterion
\[
V=\sum_{k=0}^\infty\lp \Delta v^2_k + \rho_I\Delta^2 v_{I k} + \rho_u\Delta u^2_k \rp
\]
then generates the state feedback controller 
\[
\Delta u_k =-K_P(v^r)\Delta v_k -K_I(v^r)\Delta v_{I k},
\]
where the solution of a Riccati equation \cite{anderson:2014} for the actual reference velocity $v^r$ gives the optimal gain vector $K(v^r)=\mat{K_P(v^r) & K_I(v^r)}$. This feedback gain is computed for a suitable number of reference velocities~$v^r$, depending on the size of the nonlinearity. Linear interpolation between the achieved gain vectors finally generates a velocity-dependent feedback gain for an arbitrary desired reference velocity $v^r$. 

Introducing the control error  $e=v^r-v=-\Delta v$ and the integral time constant $T_I=K_P/K_I$
results in a {\em velocity dependent PI-controller} $\Delta u_k=F_{PI}(q,v^r)e_k$ where, using the time shift operator $q$,  the corresponding transfer function is 
\beq
F_{PI}(q,v^r)=K_P(v^r)\Lp 1+\frac{1}{(q-1) T_I(v^r)}\Rp
\eeq{eq:PI}

\subsection{Feedforward and feedback control including anti-windup}
Adding the feedforward and feedback control signals generates the total control signal 
\[
u = u^r + \Delta u.
\]
Since this control signal is saturated, {\em anti-windup} is also added \cite{astrWittCCS:1997, blStudent:2002}. This means that the saturated control $u^s$, which is now the output signal from the controller, becomes 
\[
u^s = \leftp{ll}{u&|u|\leq u^{lim} \yspa{1.2ex}\\ \text{sign}(u) \sa u^{lim} & |u|> u^{lim}}
\]

\begin{figure}[t]
\b{center}
\includegraphics[height=2.75cm]{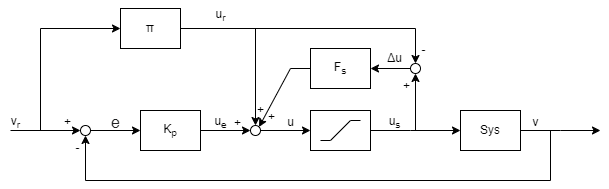}
    \caption{Feedforward and feedback control including anti-windup function.}
    \label{fig:antiwindup}
\e{center}
\end{figure}

In the anti-windup function, a low-pass filter
\[
F_s(q,v^r)=\frac{1}{1+(q-1)T_I(v^r)}
\]
is also implemented as a positive feedback around the saturation, see \figr{fig:antiwindup}. It low-pass filter the feedback control signal $\Delta u= u^s-u_r$ when the saturation is active. On the other hand, when the saturation is not active, the positive feedback loop results in the ordinary PI-controller \rf{eq:PI}, since
\[
\frac{K_P(v^r)}{1-F_s(q,v^r)} =\frac{K_P(v^r)\big(1+(q-1)T_I(v^r)\big)}{1+(q-1)T_I(v^r)-1}=F_{PI}(q,v^r)
\]

\subsubsection{Adaptive speed controller} For an arbitrary reference velocity $v^r$ we have so far obtained an adaptive speed controller. The adaptation depends on the $\theta$ vector, which is determined by estimating the parameters in the nonlinear differential equation \rf{eq:NLdiff}.

\section{Optimal reference trajectory}
The final step in the proposed modularized RL strategy is to generate an optimal reference trajectory, which describes how to move from one position to another, while for instance minimizing the energy consumption and/or the final time~$T_f$. 
The reference trajectory defines both the path to be taken and the reference velocity throughout the movement.

\subsection{Energy modeling}
In order to be able to optimize the trajectory with regard to energy consumption, an optimization objective linking the two is required. Given that 
energy is defined as power integrated over time, the objective can be formulated as 
\[
E=\int_0^{T_f} P(t) \,dt \approx \sum_{k=0}^{N-1} P_kh_k,\quad h_k=t_{k+1} -t_{k}
\]
where $P$ is the system's power output. Note that $P$ depends on the optimized reference velocity~$v^r$ and the estimated torque $u^r = \pi(v^r,\dot v^r, \alpha,\what\theta)$ as
\begin{equation} \label{eq:PowerModel}
P=C\eta(u^r)\saa u^r\saa v^r\saa,
\end{equation}
for mechanical systems, which includes a scale factor $C$ and the efficiency function
\begin{equation}\label{eq:efficiency}
\eta(u^r)=\leftp{ll}{\!\theta_7\spa{1.4ex} & u^r\geq 0 \quad \text{power generation} \\ 
\!\theta_8 &u^r<0 \quad \text{power regeneration.}}   
\end{equation}
A typical loss of $10\%$ in the mechanical power generation means that $\theta_7=1.1$, while the regeneration has about $90\%$ efficiency, implying $\theta_8=0.9$. Since $P$ is measurable and $v^r$ and $u^r$ are known, $\theta_7$ and $\theta_8$ can also be estimated by a linear regression model. Note that the scale factor $C$ is included in $P$ to account for unit mismatch, such as in the case where $u^r$ is a torque and not a force, in order to always yield power. However, since $C$ scales the entire optimization objective, its inclusion does not affect the result of the optimization. Thus, we can neglect it.

\subsection{Optimization formulation}
The optimization of the trajectory is not only dependent on the energy metric, as a multitude of constraints on the trajectory also exist. Constraints on velocity, acceleration, and control signal 
\[
0\leq v^r\leq v^{lim}, \quad |\dot v^r|\leq \dot v^{lim}, \quad|u|\leq u^{lim}
\]
are crucial to account for in the optimization. These constraints describe the requirement to stay within speed limits, obtaining comfortable accelerations, and limit the reference trajectory to only results that are feasible to follow for the system. 
Minimizing the energy $E$ while including these constraints results in an optimization problem that can be solved by any efficient optimization method, for instance convex optimization \cite{Jonas_2022_eco_drive}.

A simplification to the optimization problem is that, in many applications, the path may be considered to be known. This means that it is only the reference velocity which is required to be optimized. 
This allows for simpler optimization methods to be used, such as Temporal Optimization.



\subsection{Temporal Optimization}
The optimization method used throughout this paper is Temporal Optimization (TO). 
It assumes that the path (every position) is known, and then minimizes the energy through optimizing the sampling intervals
\[
\min_{h_0,\ldots,h_{N-1}} \sum_{k=0}^{N-1} P_kh_k, \quad\text{such that}\quad \sum_{k=0}^{N-1} h_k= T_f
\]
All time derivatives of position (velocity, acceleration, jerk) can then be approximated by difference approximations.

Adapting the formulation in~\cite{riazi2015energy} and~\cite{riazi2017energy}, the temporal optimization problem is formulated as: 
\bea
\min_{h_0,\ldots,h_{N-1}} \spa{-1.5em}&&\sum_{k=0}^{N-1}  \eta_k u^r_kv^r_kh_k \quad \text{subject to} \yspa{3.8ex}\\
  && v^r_k = (x_{k+1} - x_k)/h_k,  \nonumber\yspa{1.5ex}\\
  && \dot v^r_k = (v_{k+1} - v_k)/h_k,  \nonumber\yspa{1.6ex}\\
  && u^r_k = \pi(v^r_k,\dot v^r_k, \alpha(x_k),\what\theta),  \nonumber\yspa{1.5ex}\\
  && 0 \leq v^r_k \leq v_k^{lim},  \nonumber\yspa{1.7ex}\\
  && | \dot{v}^r_k | \leq \dot v^{lim},  \nonumber\yspa{1.5ex}\\
  && \eta_k = (1+0.1\tanh(\gamma u^r_k)), \nonumber\yspa{1.5ex}\\
  && t_{k+1} = h_k + t_{k},  \nonumber\yspa{1.5ex}\\
  && t_N \leq T_f, \label{eq:TO}\yspa{1.5ex}\\
  && t_0 = 0. \nonumber
\eea{}
where $t_k$ is the sampling time of position~$x_k$ in a sequence of~$N$ samples.
These sequences of positions and time stamps define the reference velocities $v_k^r$ and accelerations $\dot v_k^r$ that are required to follow the defined trajectory. 
Based on the reference behavior at each sample $k$, it is thus also possible to evaluate the estimated feedforward control signal to the system $u_k^r$ based on the estimated function $\pi(v^r, \dot v^r, \alpha, \hat{\theta})$. 

A benefit of TO is that the sampled positions stay the same throughout the optimization, and thus accounting for external effects or constraints that depend on positions is greatly simplified. This is useful when accounting for the slope value of sample $k$, as it will not change during the optimization given that it depends on the position $x_k$ that is unchanged. Similarly, introducing speed limits that vary throughout the trajectory is simplified in TO, as they too only depend on the sample position and thus also are constant for each sample during the optimization. 
In order to yield reasonable driving behavior for the reference trajectory, the acceleration during each segment is also constrained. 
The efficiency function of the system $\eta(u_k^r)$ is smoothed slightly from its general form given in \eqref{eq:efficiency} in order to avoid numerical instabilities in the optimization. However, the scaling factor $\gamma$ is matched to the signal range of $u_k$ such that it approximates a step function whilst ensuring numerical stability. 



One of the assumptions made here is that constant acceleration is achieved throughout each segment, and that the corresponding control signal $u_k^r$ ensures that. However, since the duration of the samples is changed in the optimization, the accuracy of this assumption will vary. This however can be mitigated by resampling the optimized trajectory with equidistant samples in time, in order to keep the segments roughly to the same time duration.

\subsection{Simplified Optimization based on Pseudo Power}
In the cases where a proper power model is unavailable, pseudo-power may instead be used in the TO to generate an approximate energy optimized trajectory. Pseudo-power is defined as acceleration multiplied by velocity. Relating this to the estimated model in equation~\eqref{eq:NLdiff}, it corresponds to $\theta_2$ through $\theta_6 = 0$, resulting in the energy criterion
\[ E = \sum_k \eta_k \theta \dot v_k v_k h_k \] 
Note here that $\theta$ is only a constant scale factor, similar to $C$ in equation~\eqref{eq:PowerModel}. Thus, $\theta$ can also be neglected in the optimization. The efficiency function $\eta_k$ is however required to include energy losses in the system, which takes the form given in equation~\eqref{eq:efficiency}, although with $u^r=\dot v_k$. The used values for the efficiencies, $\theta_7$ and $\theta_8$, will affect the optimization criterion and thus also the optimized trajectory. However, these values are constant throughout the optimization and thus can be viewed as hyperparameters for the setup of the optimization. This means that for the TO, the energy criterion from the pseudo-power formulation is only dependent on the trajectory, which defines $\dot v_k$, $v_k$ and $h_k$ for all samples $k$. 

Although pseudo-power might be seen as a simplistic model for the power estimation and by extension the benefit of optimizing against that criterion may be questioned. It was shown in \cite{riazi2017energy} that TO, using pseudo-power, produced an energy reduction of $25-30\%$ for a six-axis industrial robot while keeping the same final time $T_f$.  That paper also demonstrated a peak power reduction of up to $60\%$, demonstrating that TO using pseudo-power is a useful tool for energy optimization.

\section{Evaluation on Electric Vehicles}\label{sec:eval}
We evaluate our proposed modularized RL strategy on two electric vehicle applications. The task is to track the speed limit along a given route while using the given battery charge efficiently. In that, legal speed limits shall not be exceeded, and the total journey time must not be prolonged unreasonably. The details of the cases and our comparison to state-of-the-art deep reinforcement learning are given in the following.

\subsection{Case A: Electric Truck}
In Case A, we study the effect of the total available time~$T_f$, set in constraint~\rf{eq:TO}, on the energy consumption of an electric truck.

\subsubsection{Model} The truck model used in this paper originates from \cite{Jonas_2022_eco_drive} but with added motor dynamics. The dynamics are given by Newton's laws of motion and the feedback loop of the electric motor
\begin{align*}
    \dot{s}(t) &= v(t) \\
    m\Dot{v}(t) &= u_m(t)/R - F_{air}(v) - F_\alpha(s) \\
    T_m\dot u_m(t) &= u(t) - u_m(t) 
\end{align*}
where $m$ is the total mass of the vehicle, $t$ is the travel time, $s$ is the distance traveled, $v$ is the velocity, $u_m$ is the torque at the wheel generated by the motor, $R$ is the radius equivalent to the mechanical advantage between the motor torque and the traction force on the wheels, $F_{air}$ is the braking force generated by aerodynamic drag, $F_\alpha$ is the force generated due to the inclination of the road, $T_m$ is the closed loop motor time constant and $u$ is the control input to the system. The aerodynamic drag and impact of road inclination are modeled as
\begin{align*}
    F_{air}(v) &= \frac{\rho_a c_d A_f v^2}{2} \\
    F_\alpha(s) &= mg(sin(\alpha(s)) + c_rcos(\alpha(s)))
\end{align*}
where $\rho_a$ is the air density, $c_d$ is the aerodynamic drag coefficient,
$A_f$ is the frontal area of the vehicle, $g$ is the gravitational acceleration,
and $c_r$ is the rolling resistance coefficient. The parameter values that where used for this model in this paper are presented in Table~\ref{tab:TruckParameters}.

\subsubsection{Estimated model} A nonlinear system model $\pi_T$ of the truck was estimated, using the form structure 
\begin{equation*}
    u = \theta_1\Dot{v} + \theta_2 + \theta_3v^2  + \theta_4\alpha =\pi_T(\dot v, v, \alpha, \hat{\theta})
\end{equation*}
which is based on the base form given in equation~\eqref{eq:NLdiff}. However, initial testing yielded that both the parameters associated with the linear velocity term and quadratic slope were zero and therefore removed. Note that this model does not fully match the proper truck model, as the impact from the slope is simplified, and the motor time constant is fully neglected. The remaining parameter values were estimated to an accuracy of errors below $0.5\%$ compared to the proper model. 


\vspace{-0.3em}
\begin{table}[tb]
    \b{center}
    \caption{Truck model parameters}
    \label{tab:TruckParameters}
    \small
    \begin{tabular}{llcl}
    \toprule
         Vehicle mass & $m$ \eqqq $40\,000\,\unit{kg}$ \\
         Radius & $R$ \eqqq $0.1\, \unit{m}$\\
         Air density & $\rho_a$ \eqqq $1.29\,\unit{kg/m^3}$\\
         Aerodynamic drag coefficient & $c_d$ \eqqq $0.5$\\
         Vehicle frontal area & $A_f$ \eqqq $10\,\unit{m^2}$ \\
         Gravitational acceleration & $g$ \eqqq $9.81 \,\unit{m/s^2}$ \\
         Rolling resistance coefficient & $c_r$ \eqqq $0.006$ \\
         Closed loop  motor time constant & $T_m$ \eqqq $1\, \unit{s}$ \\ 
    \bottomrule
    \end{tabular}
    \e{center}
\end{table}

\subsubsection{Results}

\begin{figure}
    \centering
    \includegraphics[width=0.95\linewidth]{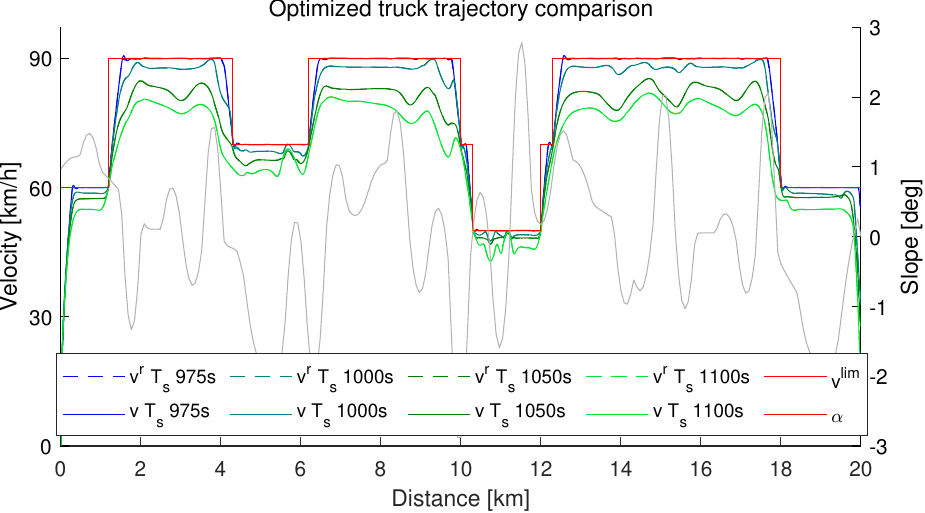}
    \caption{The graph shows the results of our method evaluated on the electric truck example, where the final time $T_f$ has been varied. The red line represents the speed limit, and the gray line the slope. The dashed lines show the reference velocities generated by the optimization, and the corresponding colored full line shows the observed behavior of the truck using that reference.}
    \label{fig:optTra_comp}
\end{figure}


The optimal reference trajectory generated by the method for the electrical truck model $\pi_T$ is shown in Fig.~\ref{fig:optTra_comp} for a driving scenario where four different total times were used. One corresponds to the minimum time for which the task is still feasible without breaking the speed limits, and three more lenient where additional time is permitted. The graph clearly shows a significant difference in optimal behavior, all different from the initial trajectory. 
As expected, the time-constrained scenario is pushed to the speed limits, while the more relaxed ones do not reach the higher speed limits. 
It is also observed that the loosened time constraint allows for greater adaptation to the slope, in order to minimize the energy consumption, where the velocity increases when the slope is negative and decreases when the slope is positive. This effect is greatest for the least constrained solution. 
In Table~\ref{tab:TruckEnergy}, the energy savings from the increase in allowed time are shown, and one can see that the relaxation of the time constraint brings a significant reduction in energy consumption.


\begin{table}[tb]
    \centering
    \caption{Truck Energy Consumption}
    \label{tab:TruckEnergy}
    \begin{tabular}{l|cccc}
    \toprule
         $T [s]$  & 975 & 1000 & 1050 & 1100\\
         slowdown & $-$ & $2.56\%$ & $7.7\%$ &  $12.8\%$  \\
    \midrule
         $\hat{E}$ & $1$ & $0.94$ & $0.91$ & $0.86$ \\
    \midrule
    \end{tabular}
\end{table}


\begin{figure}
    \centering
\b{center}
\includegraphics[width = 0.95\linewidth]{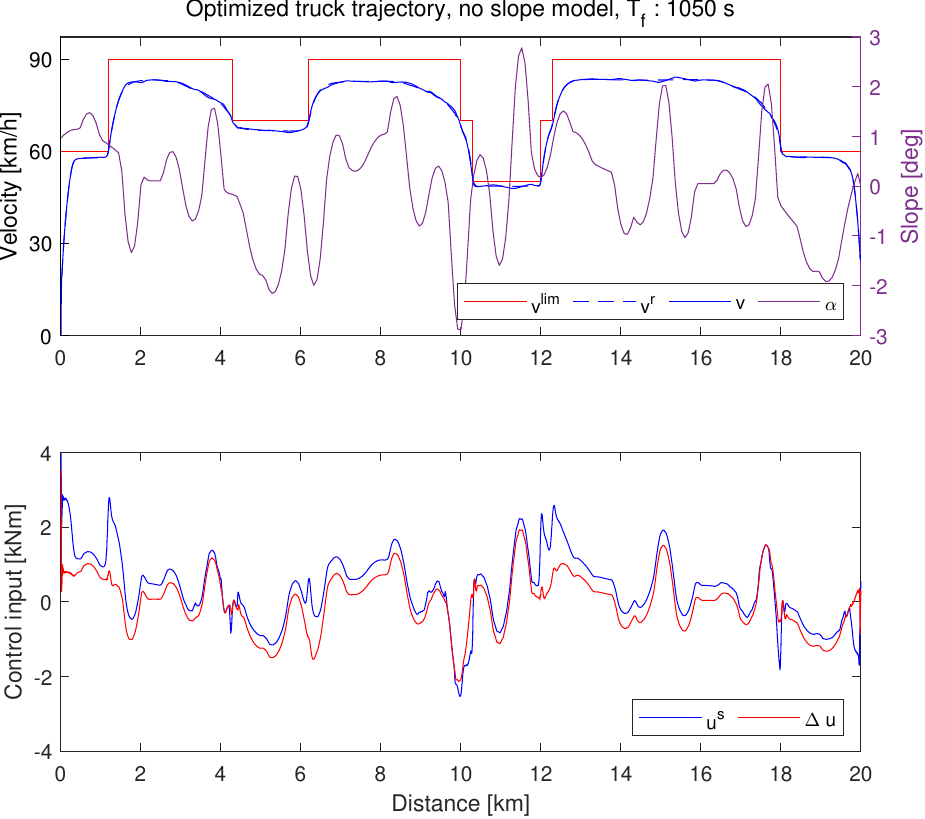}
\caption{
    The first graph shows the results of our method evaluated on the electric truck example when the estimated model did not include slope dynamics, where the dashed blue line represents the optimized reference trajectory, the solid blue line the realized trajectory, the red line the speed limit, and the purple line shows the slope of the road throughout the path. 
    The second graph shows the total control input to the system, the blue line, and the feedback control signal in the control loop, the red line. 
    }
    \label{fig:optTra_noSlope}
\e{center}

\b{center}
\includegraphics[width = 0.95\linewidth]{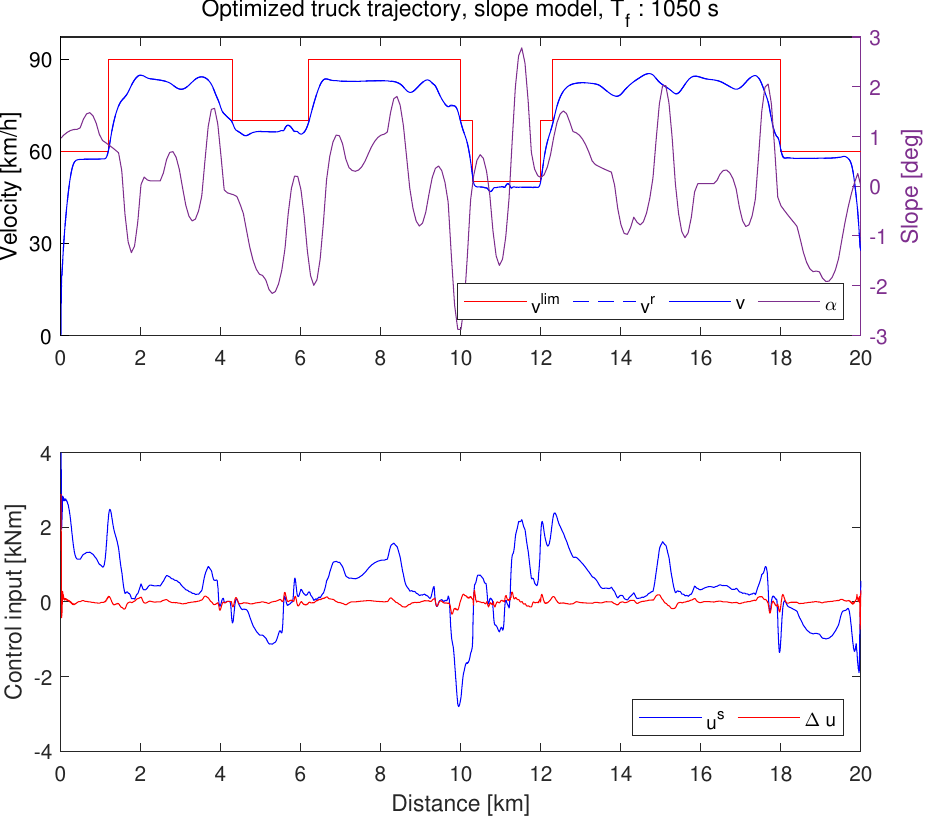}
\e{center}
    \caption{
    The first graph shows the results of our method evaluated on the electric truck example when the estimated model includes slope dynamics, 
    where the dashed blue line represents the optimized reference trajectory, the solid blue line the realized trajectory, the red line the speed limit, and the purple line shows the slope of the road throughout the path. 
    The second graph shows the control input to the system, the blue line, and the feedback signal in the control loop, the red line. 
    }
    \label{fig:optTra}
\end{figure}



\paragraph{Impact of neglecting slope}
The optimization method was also tested for the case where the slope impact was completely neglected in the estimated model of the truck. The results of this are shown in Fig.~\ref{fig:optTra_noSlope}, where it can be seen that the optimized trajectory is much smoother compared to the one generated using a model that accounted for the slope, which is shown in Fig.~\ref{fig:optTra}. In both cases, it can be seen that the system accurately tracks their respective reference trajectory, however, the controller behavior differs between the two. The controller for the model with slope included can be considered as an adaptive disturbance feedforward controller, as the model also models the load disturbance (the slope). This is concluded based on that the feedback signal $\Delta u$ is very small throughout the entire run. However, if the model was less accurate, one would expect worse performance from the feedforward control, and instead, the feedback would play a larger role. This is what is observed for the model without modeled slope impact, where the control signal instead is dominated by the feedback loop as $u^s$ and $\Delta u$ are very similar.

\subsection{Case B: BMW i3}
In Case B, we compare the performance of our proposed method to state-of-the-art deep reinforcement learning.

\subsubsection{Model} The car model used in this paper is based on the parameters of a BMW i3 electric car from 2014, presented in~\cite{dohmen21longicontrol}. The dynamics of the vehicle are described by
\begin{align}
    m\Dot{v} &= \frac{u}{v} - \frac{\rho_a c_d A_f v^2}{2} - c_rmg
\end{align}
where $m$ is the mass of the vehicle, $v$ the velocity, $u$ the power generated by the motor, $\rho_a$ is the air density, $c_d$ is the aerodynamic drag coefficient,
$A_f$ is the vehicle frontal area, $g$ is the gravitational acceleration,
and $c_r$ is the rolling resistance coefficient. The parameter values used are listed in Table~\ref{tab:BMWParameters}. The maximum power $u$ is also limited to $[-50, 75]$ \unit{kW} and the maximum acceleration $\Dot{v}$ is limited to $[-3, 3]$ \unit{m/s^2}. The power used during driving is estimated based on real test data from the Argonne National Laboratory.

\begin{table}[tb]
    \centering
    \caption{BMW i3 model parameters}
    \label{tab:BMWParameters}
    \begin{tabular}{ll}
    \toprule
         Air density & $\rho_a = 1.2 \unit{kg/m^3}$\\
         Vehicle frontal area & $A_f = 2.38\unit{m^2}$ \\
         Rolling resistance coefficient & $c_r = 0.015$ \\
         Vehicle mass & $m = 1443\unit{kg}$ \\
         Aerodynamic drag coefficient & $c_d = 0.29$\\
    \bottomrule
    \end{tabular}
\end{table}

\subsubsection{Control Objective} 
The objective for this case is to complete a track of \SI{1000}{m} within a given time, while adhering to all speed limits and minimizing energy consumed.

\subsubsection{Benchmark Methods}
To evaluate the proposed method, we compare it to two state of the art deep reinforcement learning algorithms from the stable-baselines3 python package~\cite{stable-baselines3}, namely TD3~\cite{fujimoto2018addressing} and SAC~\cite{haarnoja2018soft}. Hyperparameters of these algorithms were kept at the package defaults, since an automated hyperparameter search did not yield consistently superior results. A particular challenge in comparison to other reinforcement learning methods is including a constraint on the terminal time of the trajectory. This is easily achieved in our method, but cumbersome to be incorporated as a carefully tuned terminal reward in a reinforcement learning context. We, therefore, forgo any reward engineering and instead analyze the effect of various weights $\zeta_{Energy}$ of the continual reward component on the learned behavior.

\subsubsection{Results}
In Table~\ref{tab:sensitivity} we present a sensitivity analysis of the reward function of~\cite{dohmen21longicontrol} and compare to our method.  It can be seen in Table~\ref{tab:sensitivity} that a higher weight on energy seems to generally lead to lower energy consumption, but simultaneously increases the total time. Our method, on the other hand, achieves for a given terminal time a lower energy consumption. In particular, we set the terminal time constraint equal to the finish time of the learned TD3 policy to be able to compare the energy consumption. With this setting, our method has an energy consumption about \SI{8}{\%} lower than the DRL policy. 

\begin{table}[]
    \centering
    \caption{TD3 Results on Case B for Different Reward Weights $\zeta_{Energy}$. $\zeta_{Energy} = 0.5$ is the original setting by~\cite{dohmen21longicontrol}. $\hat{E}$ is the Estimated Energy in \SI{}{kWh}, $T$ the Total Time in \SI{}{s}.}
    \label{tab:sensitivity}
    \begin{tabular}{l|rrrrrr|r}
    \toprule
     $\zeta_{energy}$    & 0.0 & 0.1 & 0.25 & \textbf{0.5} & 1.0 & 2.0 & \textbf{Ours}  \\
    \midrule
    $\hat{E}$ & 0.209 & 0.202 & 0.187 & \textbf{0.197} & 0.172 & 0.165 & \textbf{0.181} \\
    $T$ & 77.6 & 77.2 & 85.0 & \textbf{75.2} & 87.1 & 115.3 & \textbf{75.2} \\
    \bottomrule
    \end{tabular}
\end{table}

\begin{figure}
    \centering
    \includegraphics[width=\linewidth]{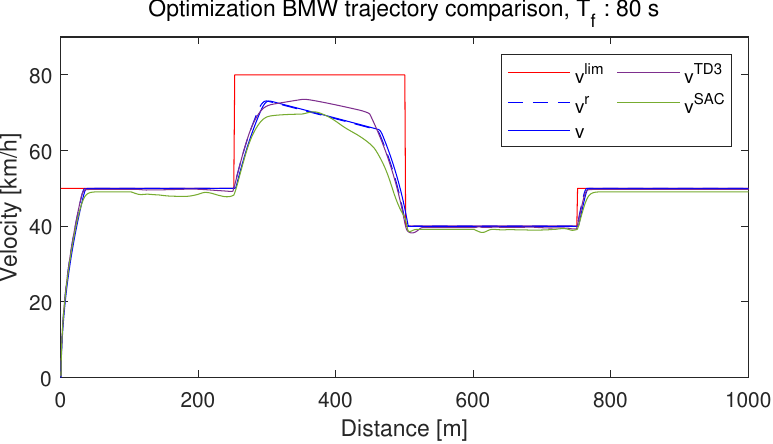}
    \caption{A comparison of our method to TD3 and SAC. Both benchmark DRL policies are the best out of 5 training repetitions spanning 500k timesteps each. All three methods arrive at relatively similar control behavior.}
    \label{fig:sarmads_method_wiering}
\end{figure}

Fig.~\ref{fig:sarmads_method_wiering} provides, in addition, policy roll-outs of our method as well as of a TD3 and SAC policy that were trained for 500k timesteps each. These deep reinforcement learning methods require on average about 50k timesteps to reach similar performance to our method. After 100k to 200k samples, TD3 and SAC start to converge on this task. Despite the strong difference in sample requirements, the three control policies do not differ significantly in behavior. The most notable difference between them is the speed around $x=400$. Here, the energy minimization leads our method to decelerate earlier than the deep reinforcement learning policies.

\section{Conclusions and topics for future research}

Motivated by robustness, data efficiency, and optimization constraints, a specific model-based RL strategy, called modularized RL, is proposed in this paper. This strategy separates the learning phase from the control optimization, meaning that the estimated model can be computed and evaluated by well-known and powerful system identification tools. In this paper, a continuous-time nonlinear state-space model is estimated. 

The control design and optimization can also be performed by powerful methods available in the control and operations research areas. In this paper, the importance of including a combined nonlinear feedforward and feedback control strategy is confirmed in an eco-driving vehicle application. The feedback loop may be designed by many alternative strategies, including simple but robust PI and PID controllers, as well as more complex robust/nonlinear feedback control strategies. In this application, the strength of a simple and flexible optimization procedure, called temporal optimization, is also demonstrated. Hard constraints are then easily introduced based on the fact that the path is known in the actual application.

Finally, we note that the model estimation is sensitive to neglected high-frequency dynamics. Sufficient accuracy was achieved, although an extra time constant was included in the heavy truck vehicle data generation, but not included in the estimated state-space model structure. Increasing this neglected time constant by one decade means, on the other hand, that the parameter errors will increase to about $10\%$. An interesting alternative is then to estimate nonlinear input-output models based on neural networks or other function approximations. The temporal optimization that has the length of individual sampling intervals as decision parameters must then be somewhat generalized. This is an interesting topic for further research.

\balance
\bibliography{references}

@inproceedings{ludvig:2025,
  title={Robust linear quadratic reinforcement learning by filtering},
  author={Ludvig Svedlund and Bengt Lennartson},
  booktitle={Proc. 21st IEEE International Conference on Automation Science and Engineering {(CASE)}},
  year={2025}
}

@article{stable-baselines3,
  author  = {Antonin Raffin and Ashley Hill and Adam Gleave and Anssi Kanervisto and Maximilian Ernestus and Noah Dormann},
  title   = {Stable-Baselines3: Reliable Reinforcement Learning Implementations},
  journal = {Journal of Machine Learning Research},
  year    = {2021},
  volume  = {22},
  number  = {268},
  pages   = {1-8},
  url     = {http://jmlr.org/papers/v22/20-1364.html}
}

@inproceedings{liu2017speed,
  title={Speed profile planning in dynamic environments via temporal optimization},
  author={Liu, Changliu and Zhan, Wei and Tomizuka, Masayoshi},
  booktitle={2017 IEEE Intelligent Vehicles Symposium (IV)},
  pages={154--159},
  year={2017},
  organization={IEEE}
}

@inproceedings{riazi2015energy,
  title={Energy optimization of multi-robot systems},
  author={Riazi, Sarmad and Bengtsson, Kristofer and Wigstr{\"o}m, Oskar and Vidarsson, Emma and Lennartson, Bengt},
  booktitle={2015 IEEE international conference on automation science and engineering {(CASE)}},
  pages={1345--1350},
  year={2015},
  organization={IEEE}
}

@article{riazi2017energy,
  title={Energy and peak power optimization of time-bounded robot trajectories},
  author={Riazi, Sarmad and Wigstr{\"o}m, Oskar and Bengtsson, Kristofer and Lennartson, Bengt},
  journal={IEEE Transactions on Automation Science and Engineering},
  volume={14},
  number={2},
  pages={646--657},
  year={2017},
  publisher={IEEE}
}

@ARTICLE{LewisRL:2012,
  author={Lewis, Frank L. and Vrabie, Draguna and Vamvoudakis, Kyriakos G.},
  journal={IEEE Control Systems Magazine}, 
  title={Reinforcement Learning and Feedback Control: Using Natural Decision Methods to Design Optimal Adaptive Controllers}, 
  year={2012},
  volume={32},
  number={6},
  pages={76-105},
  doi={10.1109/MCS.2012.2214134}}

@ARTICLE{Jonas_2022_eco_drive,
  author={Hamednia, Ahad and Sharma, Nalin Kumar and Murgovski, Nikolce and Fredriksson, Jonas},
  journal={IEEE Transactions on Intelligent Transportation Systems}, 
  title={Computationally Efficient Algorithm for Eco-Driving Over Long Look-Ahead Horizons}, 
  year={2022},
  volume={23},
  number={7},
  pages={6556-6570},
  doi={10.1109/TITS.2021.3058418}}

@article{lennartson2020reinforcement_WODES,
  title={Reinforcement learning with temporal logic constraints},
  author={Lennartson, Bengt and Jia, Qing-Shan},
  journal={IFAC-PapersOnLine},
  volume={53},
  number={4},
  pages={485--492},
  year={2020},
  publisher={Elsevier}
}

@inproceedings{cronrath2022relevant,
  title={Relevant safety falsification by automata constrained reinforcement learning},
  author={Cronrath, Constantin and Huck, Tom P and Ledermann, Christoph and Kr{\"o}ger, Torsten and Lennartson, Bengt},
  booktitle={2022 IEEE 18th International Conference on Automation Science and Engineering (CASE)},
  pages={2273--2280},
  year={2022},
  organization={IEEE}
}

@inproceedings{tu2019gap,
  title={The gap between model-based and model-free methods on the linear quadratic regulator: An asymptotic viewpoint},
  author={Tu, Stephen and Recht, Benjamin},
  booktitle={Conference on Learning Theory},
  pages={3036--3083},
  year={2019},
  organization={PMLR}
}

@book{sutton2018reinforcement,
  title={Reinforcement learning: An introduction},
  author={Sutton, Richard S and Barto, Andrew G},
  year={2018},
  publisher={MIT press}
}

@conference{dohmen21longicontrol,
  author={Dohmen, Jan and Liessner, Roman and Friebel, Christoph and Bäker, Bernard},
  title={LongiControl: A Reinforcement Learning Environment for Longitudinal Vehicle Control},
  booktitle={Proceedings of the 13th International Conference on Agents and Artificial Intelligence - Volume 2: ICAART,},
  year={2021},
  pages={1030-1037},
  publisher={SciTePress},
  organization={INSTICC},
  doi={10.5220/0010305210301037},
  isbn={978-989-758-484-8},
}

@book{bertsekas2019reinforcement,
  title={Reinforcement learning and optimal control},
  author={Bertsekas, Dimitri},
  volume={1},
  year={2019},
  publisher={Athena Scientific}
}

@inproceedings{bradtke1994adaptive,
  title={Adaptive linear quadratic control using policy iteration},
  author={Bradtke, Steven J and Ydstie, B Erik and Barto, Andrew G},
  booktitle={Proceedings of 1994 American Control Conference-ACC'94},
  volume={3},
  pages={3475--3479},
  year={1994},
  organization={IEEE}
}

@article{amodei2016concrete,
  title={Concrete problems in AI safety},
  author={Amodei, Dario and Olah, Chris and Steinhardt, Jacob and Christiano, Paul and Schulman, John and Man{\'e}, Dan},
  journal={arXiv preprint arXiv:1606.06565},
  year={2016}
}

@inproceedings{liu2021policy,
  title={Policy learning with constraints in model-free reinforcement learning: A survey},
  author={Liu, Yongshuai and Halev, Avishai and Liu, Xin},
  booktitle={Proceedings of the Thirtieth International Joint Conference on Artificial Intelligence},
  year={2021}
}

@inproceedings{fujimoto2018addressing,
  title={Addressing function approximation error in actor-critic methods},
  author={Fujimoto, Scott and Hoof, Herke and Meger, David},
  booktitle={International conference on machine learning},
  pages={1587--1596},
  year={2018},
  organization={PMLR}
}

@inproceedings{haarnoja2018soft,
  title={Soft actor-critic: Off-policy maximum entropy deep reinforcement learning with a stochastic actor},
  author={Haarnoja, Tuomas and Zhou, Aurick and Abbeel, Pieter and Levine, Sergey},
  booktitle={International conference on machine learning},
  pages={1861--1870},
  year={2018},
  organization={PMLR}
}

@book{altman1999constrained,
  title={Constrained Markov decision processes: stochastic modeling},
  author={Altman, Eitan},
  year={1999},
  publisher={Routledge}
}

@book{bds:2010,
  title={Reinforcement learning and dynamic programming using function approximators},
  author={Lucian Busoniu and Robert Babuska and Bart De Schutter and Damien Ernst},
  publisher={CRC Press},
  year={2010}
}

@book{anderson:2014,
  title={Optimal Control: Linear Quadratic Methods},
  author={Brian D. O. Anderson and John B. Moore},
  publisher={Dover Publications},
  year={2014}
}

@misc{OpenAISpinningRL,
    title = {{Spinning Up in Deep RL!}},
    author = {{Open AI}},
    url = {https://spinningup.openai.com/en/latest/}
}

@book{ljung:1999,
  title = {System Identification - Theory for the User},
  author = {Lennart Ljung},
  publisher = {Prentice Hall},
  edition = {2},
  year = {1999}
}

@manual{ljung:2025,
author = {Lennart Ljung},
title = {MATLAB System Identification Toolbox},
url = {https://se.mathworks.com/help/ident/index.html},
year = {2025}
}

@book{ljungGlad:2021,
  title = {Modeling and Identification of Dynamic Systems},
  author = {Lennart Ljung and Torkel Glad and Anders Hansson},
  publisher = {Studentlitteratur},
  edition = {2},
  year = {2021}
}

@book{astrWittCCS:1997,
  title = {Computer-Controlled Systems: Theory and Design},
  author = {Karl Johan {\AA}str{\"o}m and Bj{\"o}rn Wittenmark},
  publisher = {Prentice Hall},
  edition = {3},
  year = {1997},
}

@book{haykin:1998,
  title={Neural Networks: A Comprehensive Foundation},
  author={Simon Haykin},
  publisher={Pearson},
  edition = {2},
  year={1998}
}

@book{goodfellow:2016,
  title = {Deep Learning},
  author = {Ian Goodfellow and Yoshua Bengio and Aaron Courville},
  publisher = {MIT Press},
  year = {2016}
}

@book{blStudent:2002,
  title = {Reglerteknikens grunder},
  author = {Bengt Lennartson},
  publisher = {Studentlitteratur},
  year = {2002}
}

@article{moos2022robust,
  title={Robust reinforcement learning: A review of foundations and recent advances},
  author={Moos, Janosch and Hansel, Kay and Abdulsamad, Hany and Stark, Svenja and Clever, Debora and Peters, Jan},
  journal={Machine Learning and Knowledge Extraction},
  volume={4},
  number={1},
  pages={276--315},
  year={2022},
  publisher={MDPI}
}
\end{document}